# Overlimiting current

# in non-uniform arrays of microchannels

Hyekyung Lee[1], Shima Alizadeh[2], Tae Jin Kim[3], Seung-min Park[4], Hyongsok Tom Soh[4,5], Ali Mani[2†] and Sung Jae Kim[1,6,7†]

[1]*Department of Electrical and Computer Engineering,*
*Seoul National University, Seoul 08826, Republic of Korea*
[2]*Department of Mechanical Engineering, Stanford University, Stanford 94305, USA*
[3]*Department of Radiation Oncology, Stanford University, Stanford 94305, USA*
[4]*Department of Radiology, Stanford University, Stanford 94305, USA*
[5]*Department of Electrical Engineering, Stanford University, Stanford 94305, USA*
[6]*Nano System Institute, Seoul National University, Seoul 08826, Republic of Korea*
[7]*Inter-university Semiconductor Research Center,*
*Seoul National University, Seoul 08826, Republic of Korea*

[†]Correspondence should be addressed to Prof. Ali Mani and Prof. Sung Jae Kim
E-mail: (Ali Mani) alimani@stanford.edu, (SJKim) gates@snu.ac.kr

**ABSTRACT**

Overlimiting current (OLC) through electrolytes interfaced with perm-selective membranes has been extensively researched in recent years for understanding the fundamental mechanisms of transport and developing efficient applications from electrochemistry to sample analysis and separation. Predominant mechanisms responsible for OLC include surface conduction, convection by electro-osmotic flow, and electro-osmotic instability depending on input parameters such as surface charge and geometric constrictions. This work studies how a network of microchannels in a non-uniform array, which mimicks a natural pore configuration, can contribute to OLC. To this end, micro/nanofluidic devices are fabricated with arrays of parallel microchannels with non-uniform size distributions. All cases maintain the same surface and bulk conduction to allow probing the sensitivity only by the non-uniformity of the channels. Both experimental and theoretical current-voltage relations demonstrate that OLCs increase with increasing non-uniformity. Furthermore, the visualization of internal recirculating flows indicates that the non-uniform arrays induce flow loops across the network enhancing advective transport. These evidences confirm a new driving mechanism of OLC, inspired by natural micro/nanoporous materials with random geometric structure. Therefore, this result can advance not only the fundamental understanding of nanoelectrokinetics but also the design rule of engineering applications of electrochemical membrane.

Electrokinetics is the study of fluid and particle motions that was first reported in 1809 showing fluid migrates through the narrow spaces between naturally packed soil under the application of electric field [1, 2] (Figure 1(a)). Moving forward 200 years, advancement in understanding of these phenomena has attracted significant attention because of various technological applications in environmental [3-6], biomedical [7, 8] and energy related [9, 10] research field. Especially, the perm-selectivity of nanostructures, associated with electrical double layer overlap, has accelerated the fundamental study of nanoscale electrokinetics [11-13] and facilitated new engineering applications including heavy metal removal [14], water purification [4, 15-18], biomolecular separation [19-21] and efficient batteries [22]. In these applications, micro/nanostructures are fabricated in the form of simplified uniform geometries [23-25] to allow the precise characterization of manufacturing a system and performance of it. However, restricting the fabrication to uniform structures may result in the loss of benefits that otherwise exists due to randomness in natural micro/nanostructures. Recently developed numerical simulations [26-28] demonstrate that the non-uniformity of micro/nanostructure leads to an unexpected internal recirculating flow that has not been observed in uniform arrays of microchannels. The recirculating flow could significantly enhance ionic current because it can bring distant conductive electrolyte to the membrane and thereby open new current pathways across the ion depletion zone [29-31] formed near the membrane. The resulting excess current is directly associated with an overlimiting conductance ($\sigma_{OLC}$). $\sigma_{OLC}$, which has also been observed in regular and uniform structures,[32, 33] always requires a mechanism that promotes transport beyond the classical diffusion-drift based Nernst-Planck equation[1]. Previous studies propose that (i) surface conduction (SC) [33-35], (ii) electroosmotic flow (EOF) [12, 36, 37] and (iii) electroconvective instability (EOI) [38-40] are the main contributors of OLC as function of characteristic length scale and surface charge [33, 41]. However, all of these mechanisms consider single-compartment geometries, such as a channel or a pore without considering the possibility of network effects through multiple connected pores or channels. In an experimental study by Deng *et al* [42], the possibility of OLC enhancement due to flow loops through connected pores was hypothesized but such mechanism was not visually demonstrated. This work provides the first quantitative evidences of OLC enhancement by the critical role of channel network effects.

Here we investigate the effects using a micro/nanofluidic platform shown in Figure 1(b). Natural porous structures can be characterized by several properties including roughness, connection between pores and size distribution, *etc*. Among these, we here focus on geometric non-uniformity in order to mimic the natural pore configurations. Polydimethyl-siloxane (PDMS) is used as a building block of the device. The main and buffer channel are connected by a cation selective membrane, Nafion. See supporting information for detailed experimental procedure. They have the dimension of 1 mm (width) × 6 mm (length) × 7.5 μm (depth) and 66 of micro-fins (each fin has the dimension of 7.5 μm (width) × 2.5 mm (length) × 7.5 μm (depth)) are installed only inside the main channel. The fins are unevenly arranged to form non-uniform arrays, having different gaps ($h_w$ and $h_n$) and the wide and narrow channels are consecutively placed across the 1 mm-width main channel. The side channels in the vicinity of the membrane are used for experimental convenience [43, 44]. To keep the surface and bulk conduction as constant, $h_w$ and $h_n$ have been designed so that all configurations have the same perimeter ($l = 2(h_w+h_n+2d)$) and cross-sectional area ($A = d(h_w+h_n)$) in an unit array of microchannels as shown in Figure 1(c). In this way, we can isolate the contribution of recirculating flow from other driving mechanisms of OLC initiation. In the regimes considered here, EOI has negligible effects since all of the characteristic length scale of this device was ~ 10 μm [33, 35]. If the pore size increases over 150 μm, the major driving mechanism would be changed to EOI. Then the recirculating flow would vanish because the instability destroys an induced back flow. At the large scale the results are limited by emergence of electroconvection [45] and/or gravitational effects [46]. If the pore size is reduced to SC regime, the experimental demonstration become impossible because of extremely long experimental time and no uncharged tracer inside a microchannel of few microns. At this smaller scale the validity is limited by regimes involving double layer overlap (*i.e.* scales down to 100 nm in cross section). Also there is no induced back flow in SC regime, since all of convection is largely eliminated[35].

The first basic results are the voltage-current relation in a device filled with 1 mM KCl. The external voltage increases from 0 V to 8 V at 0.2 V/sec. As shown in Figure 2, the device has recognizable Ohmic, limiting and overlimiting current behavior as a function of the pairs of ($h_w$, $h_n$). All current values overlap in Ohmic region since the bulk conduction is kept constant. Compared to the uniform case (7.5 μm, 7.5 μm), OLCs significantly increase in non-uniform cases. Such elevations are strongly attributed to the recirculating flow.

To explain this measurement quantitatively, we consider the unit array of the non-uniform pair shown in Figure 3(a). When an electric field is applied to the domain, electro-osmotic flows [2] generate in each channel with the flow rate of

$$Q_{EO_i} = \frac{\varepsilon(-\zeta)|\mathbf{E}|h_i d}{\mu} \qquad (1)$$

where $\varepsilon$ is the dielectric constant of the electrolyte solution, $\zeta$ is zeta potential, $\mathbf{E}$ is the electric field, $h_i$ is width of the microchannel, $d$ is depth of the microchannel, and $\mu$ is the viscosity of the fluid. Also, there would be pressure build up ($\Delta P$) at the end of microchannel so that pressure-driven flow was induced to the opposite direction to the electroosmotic flow with the flow rate [47] of

$$Q_{P_i} = \frac{\Delta P}{12\mu L} h_i^3 df(h_i) \qquad (2)$$

where $f(h_i)$ can be obtained by solving the steady Stokes equations in rectangular coordinate based on high-precision numerical solutions. The numerical values of the friction factors are given in the supporting information. Specifically, the pressure gradient will adjust itself such that the recirculating flow due to pressure and electroosmosis combined in both channels becomes zero, given the dead-end condition at the membrane interface. As a result, a net recirculating flow is formed in a loop, whose flow rate is given as

$$Q_{rec} = \frac{\varepsilon(-\zeta)V_{cp}hd}{\mu L}\left[\frac{\left(\frac{h_w}{h}\right)^3 f(h_w)}{\left(\frac{h_w}{h}\right)^3 f(h_w)+\left(\frac{h_n}{h}\right)^3 f(h_n)} - \frac{h_w}{h}\right] \qquad (3)$$

where $h = 15$ μm, $d = 7.5$ μm, $h_n = h - h_w$ and $0 < h_w < 15$. $V_{cp}$ is the actual voltage applied only to the wide and narrow channels. See the supporting information for the detailed derivation of this relation as well as plots of flow rate profiles. Since the cases of $(h_w, h_n) = (7.5$ μm, $7.5$ μm$)$ as well as $(h_w, h_n) = (15$ μm, $0)$ are the uniform configurations, where $Q_{rec} = 0$, one can expect a local maximum of the recirculating flow rate in the range of $7.5 < h_w < 15$.

Using this theoretically obtained $Q_{rec}$, a numerical analysis for $\sigma_{OLC}$ is conducted and the details are given in the supporting information. Experimental $\sigma_{OLC}$ is extracted from the slope of OLC regime in I-V plot of Figure 2. Quantitative linking between theoretical $\sigma_{OLC}$ and experimental $\sigma_{OLC}$ is shown in Figure 3(b) which resembles with the recirculating flow (SI Figure 1). Since the theoretical value is calculated from a single pair and the experimental value is measured from 33 pairs of the channel, each value is normalized by max($\sigma_{OLC}$) – min($\sigma_{OLC}$). Although there is a slight difference between the theoretically obtained maximum $h_w$ (~11.6 μm) and the experimentally obtained maximum $h_w$ (~10.5 μm), it is clear that $\sigma_{OLC}$s increase with increasing non-uniformity up to a certain point and then $\sigma_{OLC}$s decrease with decreasing non-uniformity. Note that experimental $\sigma_{OLC}$ is immeasurable when $(h_w, h_n) = (>11.5$ μm, $<3.5$ μm) due to the tracer/wall interaction of visualization experiments and the fabrication limit of soft-lithography (too high aspect ratio). In addition, parabolic type flows are induced not only in single pair of channel array but also in the entire device which consisted of 33 pairs. The most-astonishing fact from the observation is that $\sigma_{OLC}$ with dimension enhances 340 % (from 0.58 nS at $(h_w, h_n)$ = (7.5 μm, 7.5 μm) to 1.99 nS at $(h_w, h_n) = (10.5$ μm, 4.5 μm)) only by slight adjusting gaps between the micro-fins.

To provide unarguable evidences of the recirculating flow, the propagation of depletion zone and the motion of tracer particles are visualized as shown in time-sequential images of Figure 4(a). The electrolyte solution of 1 mM KCl with fluorescent dye (sulforhodamine B) as a depletion zone tracer and ultra-sonicated canola oil droplets (dia.= 1~2 μm) as a flow tracer are injected to main channel. Since oil is non-polar fluid with a low surface potential so that one can minimize the electrophoretic effect [48]. See supporting information for detailed experimental methods. As a control, (i) uniform case ($h_w$ = 7.5 μm and $h_n$ = 7.5 μm) is firstly visualized. The depletion zone recedes at the same velocity in each microchannel and all of the tracers are escaped from the micro-fins, showing no internal circulation. However, slight adjustment of gaps ((ii) weak non-uniform case ($h_w$ = 8.5 μm and $h_n$ = 6.5 μm)) initiates the recirculation. First of all, the velocity difference between the receded depletion zone in narrow and wide channel begins to be developed. As a result, particles escape from wide channel reroute back into the narrow channel, confirming the recirculating flow. However, they momentarily trap in the middle of narrow channel because the weak hydraulic drag force (right to left in wide channel and left to right in narrow channel) and electrophoretic force (right to left in both channels) exerted on the particle are balanced out. Subsequent experiment with $h_w$ = 10.5 μm and $h_n$ = 4.5 μm which provides the maximum $\sigma_{OLC}$ value eliminates the trapping as shown in Figure 4(a)-(iii). Droplets coming from wide channel flow along the recirculating flow so that they return toward the membrane through the narrow channel. One should refer the supporting video for clear understanding of these visualizations. Conventional monodispersed micro-particles are unsuitable tracers in this experiment, since their high surface charge hinders the recirculating by high electrophoretic mobility and entering the ion depletion zone. Such aspect is actively utilized in bio-molecular preconcentration mechanism [11, 49] which is out scope of this work.

To support these experimental observations, theoretically calculated $Q_{rec}$s are compared with the experimentally measured $Q_{rec}$s. While the particle velocities in narrow ($u_n$) and wide ($u_w$) channel are given by $Q_{rec}/h_nd$ and $Q_{rec}/h_wd$ in the theoretical analysis, experimentally measured $u_n$ and $u_w$ should include electrophoretic motions ($u_{EPH}$) as shown in Figure 4(b). Then we have two equations ($u_n = Q_{rec} / h_nd - u_{EPH}$ and $-u_w = -Q_{rec}/h_wd - u_{EPH}$) and two unknowns ($Q_{rec}$ and $u_{EPH}$). First of all, $u_{EPH}$s obtained in this way are in the range of 4.5~12 μm/s and corresponding zeta potential of these oil droplets is in the range of -7.5 ~ -20 mV by Smoluchowski velocity equation. These values allow the droplets enter the depletion zone, which is our intention to use the oil droplet as a tracer. In addition, the range is in-line with the value of $u_{EPH}$ (-20 mV) obtained in the uniform channel case ($h_w$ = 7.5 μm).

More importantly, $Q_{rec}$s measured by these equations and theoretical $Q_{rec}$ match well as shown in Figure 4(c). As similar as plot of $\sigma_{OLC}$ (Figure 3(b)), $Q_{rec}$ also has a similar trend. Although there is a slight difference between the theoretically obtained maximum $h_w$ and the experimentally obtained maximum $h_w$, $Q_{rec}$ increases with increasing non-uniformity up to a certain point and then $Q_{rec}$ decreases with decreasing non-uniformity. Note that the values of $Q_{rec}$ in Figure 4(c) are dimensionalized form because experimental $Q_{rec}$s are measured in one pair of non-uniform channel.

Based on the I-V measurements (Figure 2) and the flow visualizations (Figure 4(a)), we can confirm that the recirculating flow from induced-pressure let additional ion carrier transport through the membrane so that $\sigma_{OLC}$ significantly enhances. Consequently, we prove geometric non-uniformity causing recirculating flow as a new mechanism driving OLC in configurations involving perm-selective ion transport. In contrast to the EOI mechanism, where additional current paths are formed randomly[50-56], here the paths are regularly generated with flow in narrow channels always moving towards the membrane and in wide channels moving away from it. Thus, the present arrangements of microchannels would enhance $\sigma_{OLC}$ in more controllable manner than that of EOI. In addition, one may extract a scaling law between geometric non-uniformity and $\sigma_{OLC}$, if one can setup a proper measure of the non-uniformity. One example of modifying non-uniformity is implemented by fabricating repeated wide or narrow channels indicated as *w* and *n* shown in Figure 4(d). As demonstrated, uniform case (…$h_{7.5}h_{7.5}h_{7.5}$…) has the lowest $\sigma_{OLC}$, while strong non-uniform case (…*wnwn*…) has the highest $\sigma_{OLC}$. As the non-uniformity decreases in repeated configuration (*i.e.* …*wwwnnn*… or …*wwwwwnnnnn*…), $\sigma_{OLC}$ is significantly reduced because either *nn* pairs or *ww* pairs contribute as uniform arrays. Note that the bulk and surface conduction are kept constant to allow proper comparisons in all cases. See the supporting video for the propagation of depletion zone in each configuration.

In this work, we experimentally demonstrate perm-selective ion transportations with non-uniformly distributed microchannel network that closely mimicks a natural pore configuration. Micro/nanofluidic devices are fabricated with arrays of parallel microchannels with either uniform or non-uniform size distributions. The setups where designed such the geometric non-uniformity could be the only factor affecting OLC. All of evidences presented here including theoretical analysis, I-V measurements and flow visualization, point that natural configuration (*i.e.* non-uniform sized pores) enhances $\sigma_{OLC}$ due to their inherent non-uniformity. Compared to the previously reported mechanisms of OLC (SC, EOF, EOI, and diffusioosmosis [57]). the newly identified mechanism is unique in the sense that it requires a network of connected geometries and is characterized by geometric non-uniformities in such networks. The presenting work complements more general theoretical model of geometric non-uniformity that include effects such as complex connections, finite EDL thickness effects, and diffusion-osmosis[28] by demonstrating a direct measurement of internally induced flow loops and quantification of its impact on transport. Thus, this result can advance not only the fundamental understanding of nanoelectrokinetics but also the design rule of engineering applications of electrochemical membrane.

**ACKNOWLEDGEMENTS**

This work is supported by Basic Research Laboratory Project (NRF-2018R1A4A1022513) and the Center for Integrated Smart Sensor funded as Global Frontier Project (CISS- 2011-0031870) by the Ministry of Science, ICT & Future Planning and Korean Health Technology RND project, Ministry of Health and Welfare Republic of Korea (HI13C1468, HI14C0559). Also this work is partially supported by BK21 plus program at Seoul National University. Professor Ali Mani is financially supported by the U.S. National Science Foundation under Award# 1553275. Professor SJ Kim acknowledged the financial support from LG Yonam foundation.

**FIGURE CAPTIONS**

**Figure 1** Schematic diagram of electrokinetic flows generated by perm-selective ion transport in (a) natural micro/nanopore configuration and engineered configuration mimicked natural pores. $V_{app}$ is the external voltage to the entire channel. (b) Image of fabricated micro/nanofluidic device and (c) the combination of gaps ($h_w$, $h_n$) between micro-fins.

**Figure 2** Current-Voltage plots in non-uniform arrays of microchannels.

**Figure 3** (a) Schematic diagram of electrokinetic flows in a single non-uniform pair of microchannels. $\phi$ is the electrical potential. (b) Plots of OLC extracted from Figure 2 and recirculating flow rate by theoretical calculation as a function of $h_w$.

**Figure 4** (a) Visualizations of electrokinetic flow (particle motion and propagation of depletion zone) for uniform, weak non-uniform and strong non-uniform case. (b) Diagram of experimentally measured velocity components in each channel. (c) Plot of experimental and theoretical $Q_{rec}$s as a function of $h_w$. (d) Depletion zone propagation and experimentally measured $\sigma_{OLC}$s in repeated wide or narrow channels.

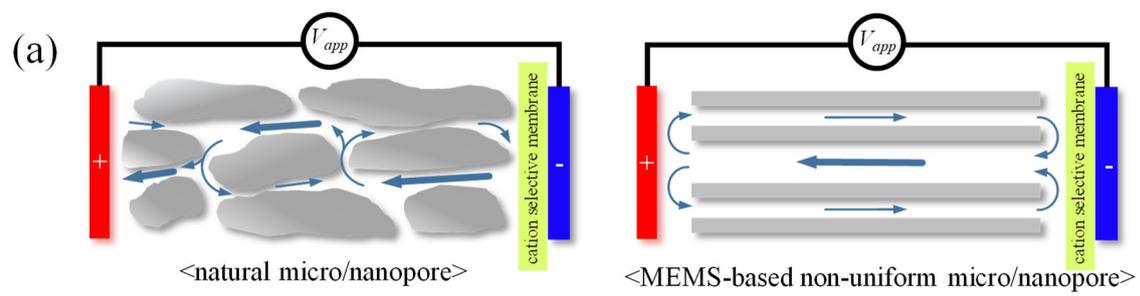

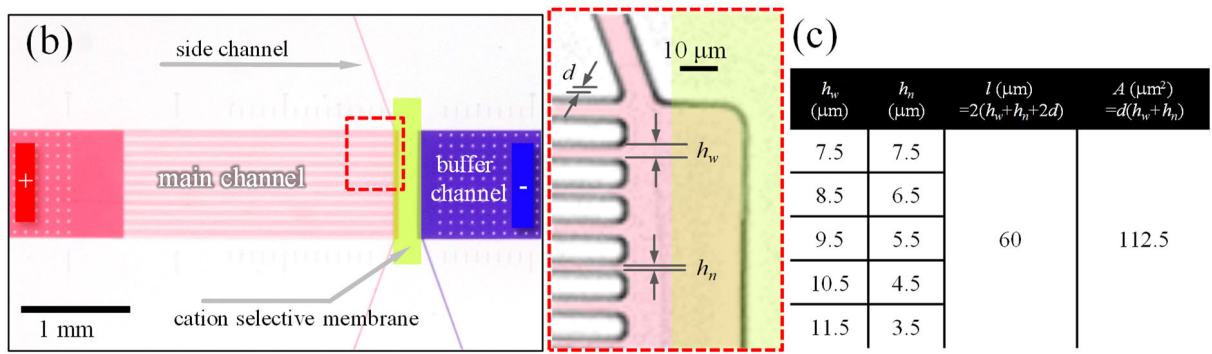

Figure 1

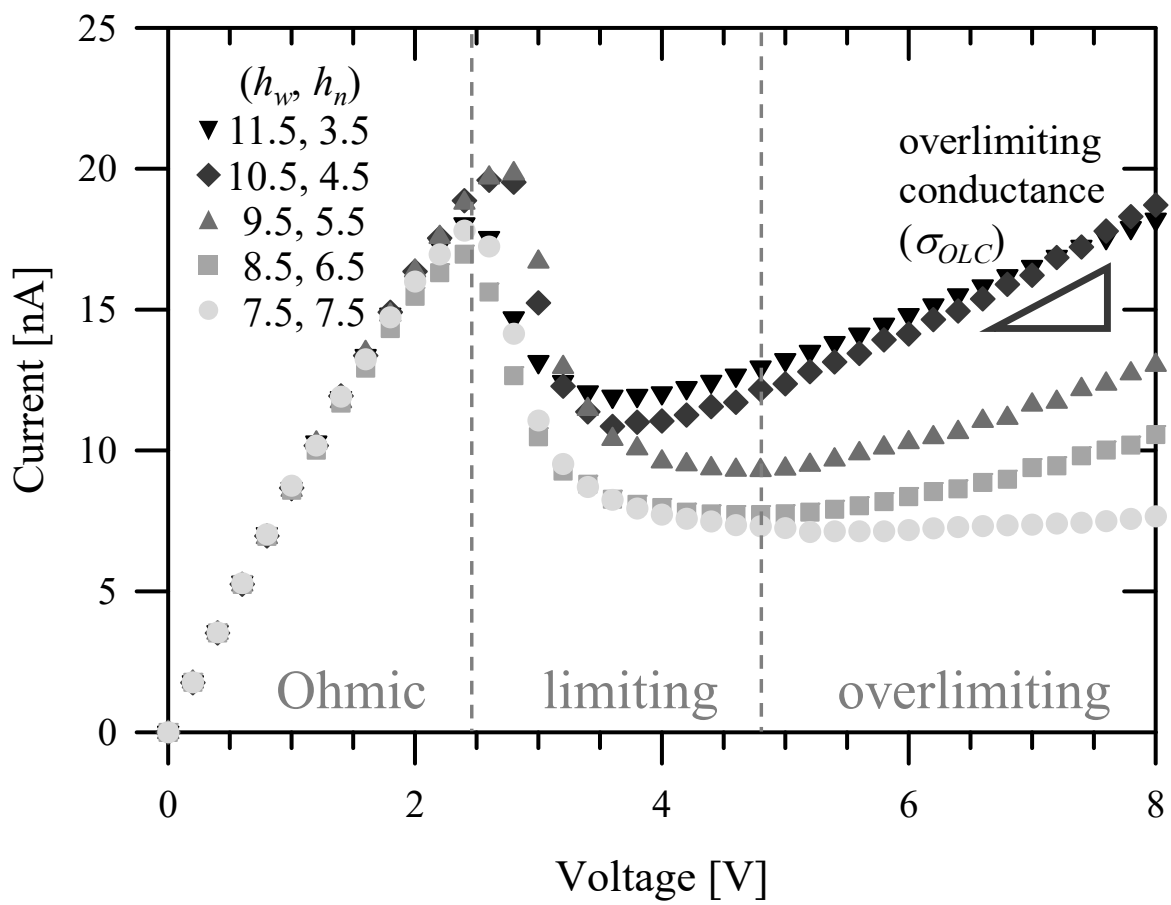

Figure 2

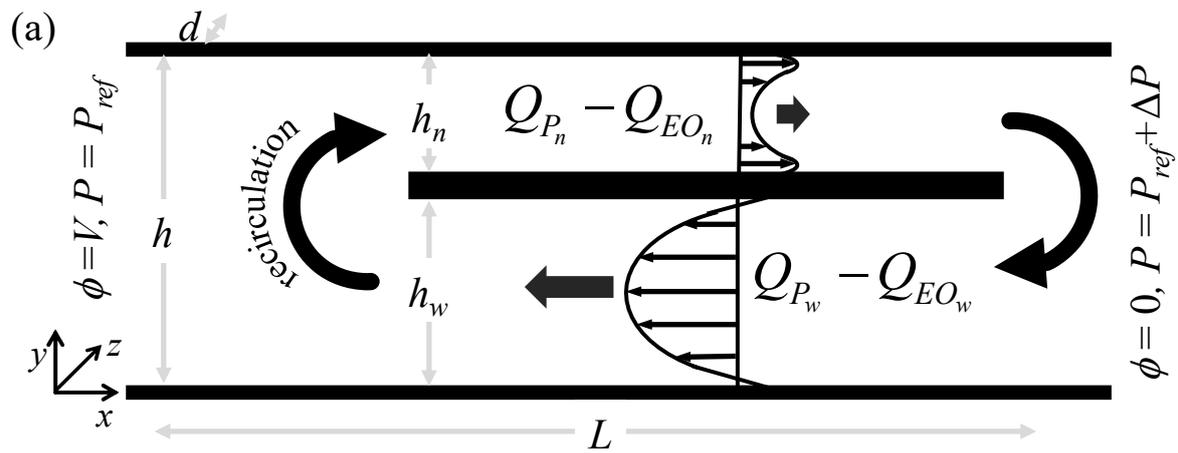

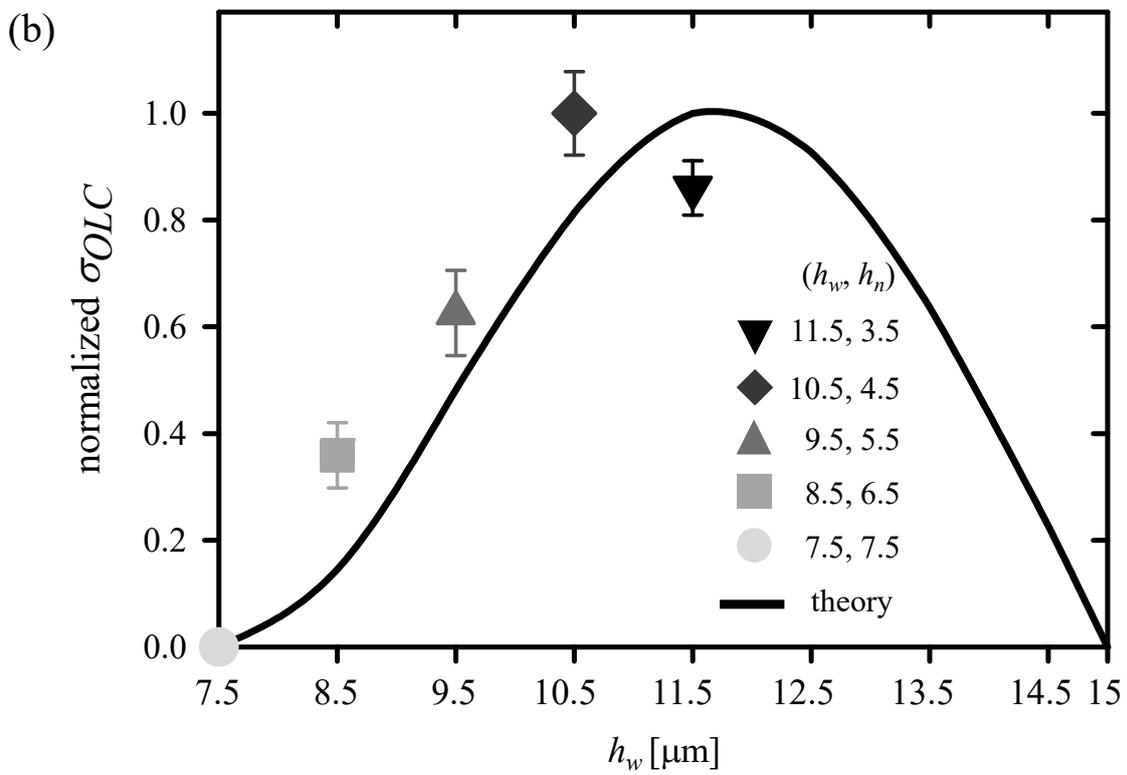

Figure 3

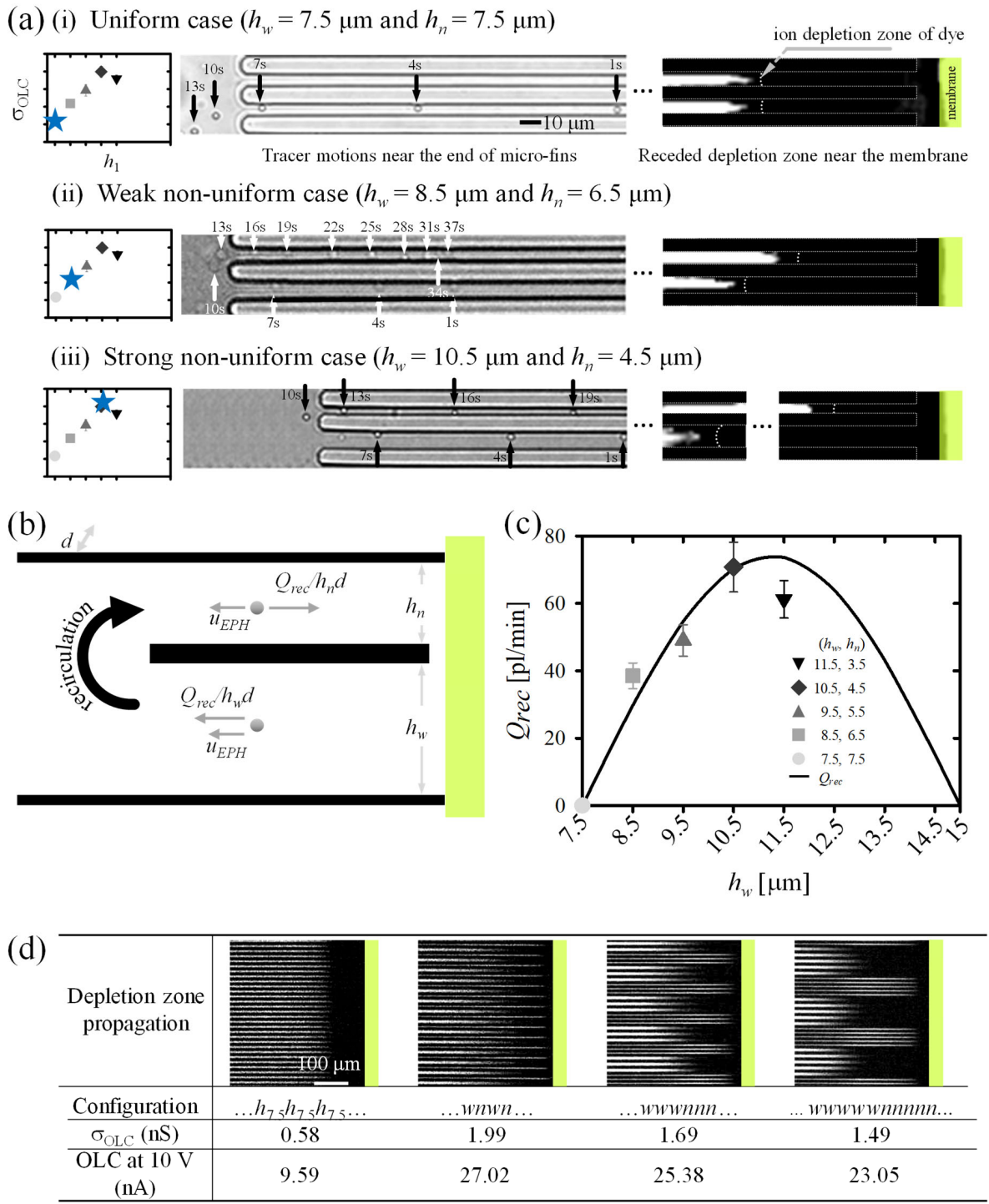

Figure 4